\documentclass[11pt]{amsart}

\usepackage{graphicx}
\usepackage{amssymb}
\usepackage{hyperref}
\usepackage{float}
\usepackage{eucal}
\usepackage{lineno}
\usepackage{amsaddr}
 
\numberwithin{equation}{section}

\usepackage{xcolor}

\title[Kaplan-Meier type curves for COVID-19]{Kaplan-Meier type survival curves for COVID-19: a health data based decision-making tool}

\author[J.M. Calabuig,  L.M. Garc\'{\i}a-Raffi, A. Garc\'{\i}a-Valiente and E.A. S\'anchez-P\'erez]{J.M. Calabuig$^{1}$, L.M. Garc\'{\i}a-Raffi$^{1,\star}$, A. Garc\'{\i}a-Valiente$^{2}$ and E.A. S\'anchez-P\'erez$^{1}$}

\address{$^{1}$
Instituto Universitario de Matem\'atica Pura y Aplicada. Universitat Polit\`ecnica de Val\`encia. Camino de Vera s/n, 46022 Valencia. Spain}

\address{$^{2}$ Universitat de Val\`encia. Burjassot 46100, Val\`encia. Spain
}
\email{jmcalabu@mat.upv.es, 
lmgarcia@mat.upv.es, algarva5@alumni.uv.es, easancpe@mat.upv.es}

\begin{document}
\maketitle

\begin{abstract}
Countries are recording health information on the global spread of COVID-19 using different methods, sometimes changing the rules after a few days. They are all  publishing the number of new individuals infected, recovered and dead, along with some supplementary data. These figures are often recorded in a non-uniform manner and do not match the standard definitions of  these variables. However, using data from the first wave of the epidemic---february-june---, in this paper we show that the Kaplan-Meier curves calculated with them could provide useful information about the dynamics of the disease in different countries.  We present a robust and simple model to show certain characteristics of the evolution of the dynamic process, showing that the differences of evolution among the countries is reflected in the corresponding Kaplan-Meier-type curves. We compare the curves obtained for the most affected countries so far, with the corresponding interpretation of the properties that distinguish them.
\end{abstract}



\section{Introduction}

Since its first detection in China, COVID 19---disease caused by SARS-CoV-2 virus---has spread to different parts of the world until reach the category of pandemic in a short period of time. This has created a social and scientific challenge, where understanding how the virus behaves is crucial in order to stop its spread. 
A classic tool in the analysis of epidemics that could be used in this sense is the Kaplan-Meier (KM) survival model  \cite{KM} which allows to calculate the step-by-step survival probability of a fixed group of patients suffering from a disease (see for example \cite[S.15]{Ji} for a contextualised explanation of the topic). 
 We propose an easy method to compute this probability \cite{cal}, which would allow the associated Kaplan-Meier-type curve to be predicted: that is, a prediction of how, given an average infected individual, his or her condition with respect to the infection changes over time.
However, mainly due to the diverse nature of data collected on the pandemic in different countries, this model needs to be adapted to the specific case of COVID-19 to provide relevant information.  As will be shown in the paper, each country has its own survival curve with strong differences, which cannot be justified as a unique consequence of local population characteristics: a survival curve should only depend on the virus, assuming the usual degree of homogeneity in the infected population. Therefore, the reason for the strong difference in the results in different regions have to be sought in two directions: first, the way countries are reacting to the epidemic, and second, the characteristics of the data these countries have made public. 


In addition, the actual models are not sensitive enough to capture the different dynamics that different strains of the virus have. This happens because viruses having RNA as their genetic material are less stable than those having DNA and tends to accumulate a greater number of mutations. This means that the virus changes more quickly, ending up in different variants of the same virus that have different mortality and infection rates. Furthermore this feature raises the fear about future cases of re-infection, where the virus differs enough from previous versions to evade the immune system again---as such other virus with the same genetic material do, e.g. Influenzavirus A, that causes the common flu and is able to infect us repeatedly \cite{Do}---. This scenario may occur as other coronaviruses are able to infect humans periodically such as HCoV-NL63 or HCoV-229, that are responsible of one out of five colds \cite{DT}. The mutation rate for SARS-CoV-2 is not known yet, but given its potential, the possibility should be considered.  
 
Thus, although all these arguments could influence the unusual results of the survival curves these facts do not substantially change the structure of the model \cite{cal}. Therefore, the problem comes from the data. 
But this fact does not invalidate the usefulness of the Kaplan-Meier curves. Here, we show that some significant patterns can be detected by comparing the curves constructed for different countries. In further applications, survival curves could also provide some useful information for decision-making regarding the application of strategies against COVID-19 spread, such as the duration of confinement periods or the intensity of policies to detect new cases.

In this work  we use the available data of the dynamics of the disease COVID-19 to understand the survival of the virus that causes it, SARS-CoV-2. Although more information is already available on the second wave of COVID-19 in the countries we have analyzed, we have chosen the methodological approach of using data from the first wave---February to June---because this period defines a complete (almost closed) cycle of infection. Since we are interested in drawing some methodological conclusions from the experience, we believe that this procedure allows a more stable framework for obtaining them. 

 The results of our analysis are the estimates of the probability distributions of virus survival in different countries. These are functions that are sensitive to changes in the epidemiological data of different populations, making the model adaptable to reinfection scenarios and other more subtle differences such as the virulence of different strains \cite{Ta}. Together with some usual models for predicting the amount of new infected population, this allows the development of a complete model for the evolution of new infected individuals, people who must be kept in quarantine and individuals who have already overcome the disease. To approximate the solution of the equations we use a genetic algorithm approach \cite{Yu}, which provides estimates of the probability and therefore clear images of the expected infection scenario.  A full explanation of the mathematical method we have developed for this is available in \cite{cal};  in the present work we show what may be relevant for the management of health systems, that is, the final results for the different countries. We believe that these results can make it possible to control the effectiveness of containment policies, thus helping in the decision-making process. The simplicity, both of the model itself and of its calculation and interpretation, is one of the main advantages of our approach, which makes it suitable as a forecasting tool.

Regarding  other models being used in the pandemic crisis, a great effort is being made to improve the mathematical representation of the number of newly infected individuals in order to provide an accurate predictive tool.  The most popular model being used is the SIR model and modifications of this model, that in particular provides a forecast of the number of new infected people in subsequent steps of the dynamical process (see for example \cite{cho,Ji} and the references therein). However, the probability of survival of the virus could be even more relevant for the management of strategical information for
decision-making concerning important data  that affects the population in  different countries. For example, to decide how long a period of confinement should last and to what type of population it should apply.

\section{Methods}

Our model  is based on a modification of the classical Kaplan-Meier survival curve. The idea is to fit the evolution curve of the accumulated total amount of {\it recovered} ($\mathcal R$) plus {\it dead} ($\mathcal D$) people---data provided by countries---from the beginning of the epidemic to time $t$. We call this number $\mathcal X$ and we will refer it as {\it discharged people.} We consider  $\mathcal X=\mathcal R+\mathcal D$.  On the basis of the accumulated total amount of {\it infected} people, ($\mathcal  J$), in the same period---data provided also by countries---,  the model fits $\mathcal J$ versus the discharged ones ($\mathcal X$) estimating the step-by-step probability of the {\it virus to survive}---denoted by $\mathcal P$---in a given infected patient. The result of our fit provides the time series of both the probabilities of survival of the virus, $\mathcal P$, and the approximation $\widehat{\mathcal X}$ of the accumulated number of discharged people ($\mathcal X$).
 
On the one hand,  in the right panel of Figures \ref{fig:survivalUSAUKSW}, \ref{fig:survivalCHKOGE}  and  \ref{fig:survivalSPITFR}   it can be seen examples of the
approximation of the function $\mathcal X$ for different countries. The black line corresponds to the real data of discharged people ($\mathcal X$)
while the red line is the result of our curve fitting ($\widehat{\mathcal X}$). As it can be seen, both curves are almost coincident for all the countries considered and in a period of time  of $95$ days. The reddish shaded area represents a range of $10\%$ over the maximum value of $\mathcal X$.

On the other hand, in the left panel of Figures \ref{fig:survivalUSAUKSW}, \ref{fig:survivalCHKOGE}  and  \ref{fig:survivalSPITFR} 
we show the representation of the survival curve of the virus. It gives the probability of an individual continuing to be infected---in terms of being under the control of the national health system according to the data collection in each country---after the day  when he/she was {\it labelled as  infected} (which corresponds to $t=0$ in the representation). 

The Kaplan-Meier (KM) survival curve \cite{KM} is based on the estimation of the instantaneous probability of survival at a given time in the process of reduction of a given population. The interested reader can find a complete explanation of this and related topics in \cite[Ch.2]{kle} and  \cite{bai,kee}. We assume that the time variable has discrete values. For the sake of simplicity of formulas and without loss of generality  we will consider  $t \in \mathbb N$ starting at the moment $t=0.$ We write $\mathcal P(t)$ for the probability of an individual that has been labeled as ``infected" to continue being infected by the virus at the time $t.$ An estimate of this value for a given population of $N$ infected individuals at the time $t=0$ is given by
$$
\mathcal P(t)= \frac{n(t)}{N}, \quad t =0,1,2,3,...,
$$
where $n(t)$ is the number of patients that are still infected at the day $t$. Note that $n(0)=N$ and $\mathcal P(0)=1.$

 Let us center now our attention in the process of infection that started in the first wave of the epidemic. Given a fixed country, 
 let us write now $I: \mathbb N \to \mathbb N$ for the function that gives the number of new infected individuals $I(t)$  at time $t.$ 
The total amount $\widehat{\mathcal X}(t)$ of individuals surviving after a time $t$ can be written 
as a KM type survival function given by the convolution formula
$$
\widehat{\mathcal X}(t)=\sum_{s=0}^t     I(s)\cdot \big( 1- \mathcal  P(t-s) \big), \quad  t =1,2,3,...,
$$
where $\mathcal P(u)$ is, as we said, the probability  that an individual will continue to be infected at the time $u.$ 
This quantity approximates the number of discharged people $\mathcal X(t).$ Using this equation it is also possible to compute the Kaplan-Meier survival curve $\mathcal P,$ starting from the data extracted of the reports of the different countries affected by the pandemia. The complete explanation of how this can be done can be found in \cite{cal}.  

\section{Results and discussion}

In the previous section we have introduce a model based on variables that are directly related to the epidemiological data provided by the states during the first wave of the Covid-19 pandemic. The meaning of the quantities appearing in this model has a direct interpretation. Let us start with the explanation of the probability distribution $\mathcal P.$
For example, in the case of Spain (Figure  \ref{fig:survivalSpain}) $62$ days after being classified as infected a standard patient has a probability of remaining infected of $0.2$. Or, in other words, $20\%$ of the patients listed as infected the first day ($t=0$) will continue labeled as infected after $62$ days (that could stay at hospital or at home in quarantine). It can be seen that there is a significant decrease in the curve in the first $10$ days. Indeed, since $\mathcal S(1)=0.96$  then  after one day $96\%$ of infected people will remain infected whereas $10$ days after this will be the case only for the $53\%$ of the infected people. However we need $52$ days more to reduce the percentage to $20\%$. The size of the balls that make up the curve is proportional to their value at the point.  The structure of these curves is the main element of our analysis.

Here, the most remarkable feature is that, as can be seen in Figures \ref{fig:survivalUSAUKSW}, \ref{fig:survivalCHKOGE} and \ref{fig:survivalSPITFR}, the model is sensitive to the progression of the epidemic in different countries, showing different patterns of survival curves. 

In countries as the United States or the United Kingdom, the form of the curves suggests that the spread of SARS-CoV-2  was not been effectively controlled at an early stage,  either because no general testing of infected people was done or the government health responsibles decided to present the global numbers in a different way,  not  counting a big group of people suspicious of being infected. As a consequence, the reported number of  'admissions' in the system (registered infected people  $\mathcal J$) is greater than the number of discharges ($\mathcal X$) over a long period of time, so it takes longer to reach equilibrium. This translates graphically in an individual's probability of getting out of the group of infected people that decreases slowly (Figure \ref{fig:survivalUSAUKSW}).

At the other end, there are countries where, after the first cases were detected, mobility was restricted and a large number of tests were carried out to identify and isolate infected persons---as the case of South Korea or Germany---. The infection counts reported by these countries  reveals this fact.  The curves suggest that this policy was maintained throughout the entire process of the first wave of the epidemic. The number  'admissions' , although initially much higher, is rapidly decreasing, approaching the number of 'discharges'. The graph shows a rapid decrease in the probability of an individual remaining infected, followed by a flattening of the curve in which a slower decrease is observed corresponding to the normal evolution of infected individuals in hospitals (Figure \ref{fig:survivalCHKOGE}). This would also be caused by---or together to---, a powerful campaigne of test made over all the general population, getting and reporting---and including in the counts--- a big number of individuals that were positive but asymptomatic, or which had a very good response to the medical treatment. 
In some cases such as in Korea, since the number of infected persons is not so great, the model shows the changes in the trend with greater sensitivity. This allows us to see how the initial trend is similar to that observed in countries with late, deficient or ineffective control measures, with a strong decrease immediately afterwards.

Finally, in countries such as Spain or Italy (Figure \ref{fig:survivalSPITFR}), where the measures taken have partially slowed down the expansion, a less pronounced decline in the KM curve is observed, exhibiting a mixed behaviour between the two extreme cases that have been considered in Figures  \ref{fig:survivalUSAUKSW} and \ref{fig:survivalCHKOGE}.

 Thus, the KM survival curve gives an idea of the speed of the national system to detect and manage new infected individuals (Figure \ref{fig:survivalpattern}).  A large number of tests allows to control a relevant number of infected individuals (perhaps asymptomatic) reducing the stress for the national health system because can reduce  the severity of the infections, i.e.,  the period infected people is under control of the health system (with a lesser use of clinical resources). This results in a considerable efficiency of the system, mainly if done at an early stage of the epidemic, and (looking at the results) seems to be the most effective strategy. Early detection (at any stage of the process, but mainly at the beginning) and  massively testing, along with containment measures to reduce the rate of infection once it has begun, appear to be the main weapons against the virus. Containment also appears to be an effective tool, but its effectiveness is based on other aspects of the system: it clearly reduces the number of new infections, but this may not affect the survival curve.


In short, we can consider that the model can help the decision-makers of each country to understand the distribution of time periods in which  the public health system has to take care of  infected people, according to the same  variables that the health policy makers have chosen to measure, in our case, infected (confirmed), recovered and dead people. Finally, in Figure \ref{fig:survivalpattern507090_USA_Korea_Spain} we show the variation of the survival curve of the virus,  when computed with different  time series of days ($50$, $70$ and $90$ days) that provide an idea of the stability of the solutions. Note that the principal feature of the curve, the decreasing in probability during the early period is maintained independent of the number of days considered.

The values of the slopes of the final parts of the survival curves are shown in Table \ref{table1}. Countries that show a strong decrease after a few days, at the beginning of the curve, have lower values of the slope. This could be attributed to better medical treatment in countries with steeper slopes, as the proportion of people leaving the system increases. However, as we assume that the medical methods used in all the countries analysed are similar, this interpretation does not seem to be correct. Instead, it seems to be a consequence of the higher proportion of the population analysed in the countries with small slopes: more people start to be followed by the health systems at an early stage in countries such as South Korea or Germany, so the medical prognosis is statistically better. Therefore, of the total population followed in these countries, only a small rate needs medical attention in the late stage, but these patients need it for a long time, so the slope is small.

\begin{table}[h!]

\begin{center}
\resizebox{.95\textwidth}{!}{
 \begin{tabular}{c c c} 
 \hline
 USA  & United Kingdom & Sweden \\ [0.5ex] 
 \hline 
-0.0106  & -0.0102 & -0.0109 \\ 
 \hline
\end{tabular}
\quad
\begin{tabular}{c c c } 
 \hline
 China & South Korea & Germany  \\ [0.5ex] 
 \hline 
 -0.0025   &   -0.003 &  -0.003  \\ 
 \hline
\end{tabular}
\quad
\begin{tabular}{c c c} 
 \hline
  France & Italy & Spain \\ [0.5ex] 
 \hline 
 -0.008 & -0.0087 & -0.006 \\ 
 \hline
\end{tabular}}
\end{center}
\caption{Slopes of the linear final trends of the KM curves. } \label{table1}
\end{table}

The data used in this work have been collected from the {\sf Github} of  the ``COVID-19 Data Repository by the Center for Systems Science and Engineering (CSSE) at Johns Hopkins University'' accessible through \url{https://github.com/CSSEGISandData/COVID-19}.  More precisely, we have made use of the time series of confirmed, dead and recovered people available through the link  \url{https://github.com/CSSEGISandData/COVID-19/tree/master/csse_covid_19_data/csse_covid_19_time_series}.

%
%
%
%
%

\section{Conclusions}

We have presented  the estimates of the probability  functions for the virus that have been computed using the data available for the first wave in nine countries, that in a sense represent three different ways of data collection and health system managements. 
It has been observed by epidemiology experts, data scientists and in general the whole public that the counting of  new  infected, recovered and dead people depends on the country---the tools are not at all homogeneous---and do not reflect the actual situation mainly with regard to new cases of infected persons. We have found the probability function for each country with the information made public by the corresponding governments during the first wave of the pandemic, because it reflects the parameters that those same governments are capable of measuring and on which they can base their strategies.



The main characteristic of the curves is possibly the initial behavior, which allows us to group the nine countries we have selected into three categories.
 In our interpretation, this initial behaviour reflects the way in which national health systems are measuring the whole of the infected population: how many patients with some symptoms have been tested, and how they decide whether they should be under the supervision of the national health system or not. 
Despite  the well-known fact that data offered by the different countries are deficient, it is precisely the difference on the shape of the curve among the countries that makes it a useful tool from the point of view of epidemiology and the management of public health systems. Under this assumption we have shown that the curve allows to group countries depending on the strategy followed  to deal with the pandemic desease, and in consecuence it contains useful information about the different actions taken. 
Regardless of how the variables are defined in each country---this has to be taken into account by the country itself when interpreting the results---the KM curve shows how quickly the public health system is able to deal with infected individuals:
the faster the decrease of the KM curve in the first steps, the less pressure the system has to bear, since individuals need to spend less time controlled by this system. We stress that this control is a matter of how each country measures infection, and must be understood in the context of each country. Different regions within a country could follow the same rules, and so could be compared.

However, some general conclusions can be drawn. The main one is that extensive population-based testing of COVID-19 improves the overall rating of the efficiency of the public health system. This is clearly demonstrated by the survival curves in Germany and Korea, compared to other countries. Since this can help control infected people, it allows countries to manage the health system resulting in a rapid decrease in the KM curve.


Finally, let us recall that the KM curve does not give a measure of how good are medical treatments to the infected people in each country. Instead,  once the counting method is fixed in each country, the KM curve provides decision-makers with a strategic tool for that country, as it gives a clear idea of how much time the health system has to take care of an infected individual, whatever this means in the particular country's statistics. This could be relevant, for example, for the installation of emergency hospitals, the duration of special confinement measures, and other extraordinary measures.

{\bf Acknowledgements.} The authors has been supported by the C\'atedra de Transparencia y Gesti\'on de Datos, UPV-GVA, Generalitat Valenciana.\\

{\bf Competing Interests.} The authors declare that they have no competing financial interests.\\

{\bf Materials \& Correspondence.} Correspondence and requests for materials should be addressed to L. M. Garc\'{i}a-Raffi (email: lmgarcia@mat.upv.es).\\
 
{\bf Supplementary information.} Is available for this paper in the file Supplementary material. 

\newpage 

\begin{center}\scriptsize
\begin{figure}[H]  
\centering {Survival curve of the virus $\mathcal S$ corresponding to Spain.}\\[2mm]
\centering\begin{minipage}{\textwidth}
\includegraphics[width=\textwidth]{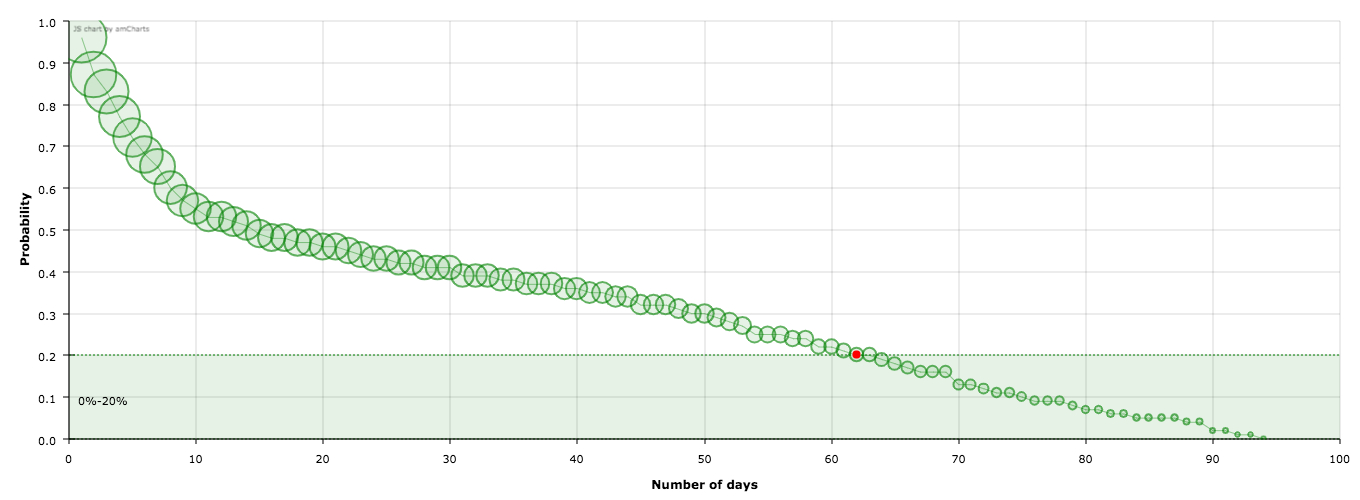}\\
\end{minipage}\\[2mm]
\caption{Survival curve, $\mathcal S$, of the virus corresponding to Spain. The red point signs the number of days after which a standard individual has a probability of staying in the group of infected people smaller than $0.2$.  \label{fig:survivalSpain}
} 
\end{figure}

\begin{figure}[H] 
\begin{minipage}{.99\textwidth}
\includegraphics[width=.5\textwidth]{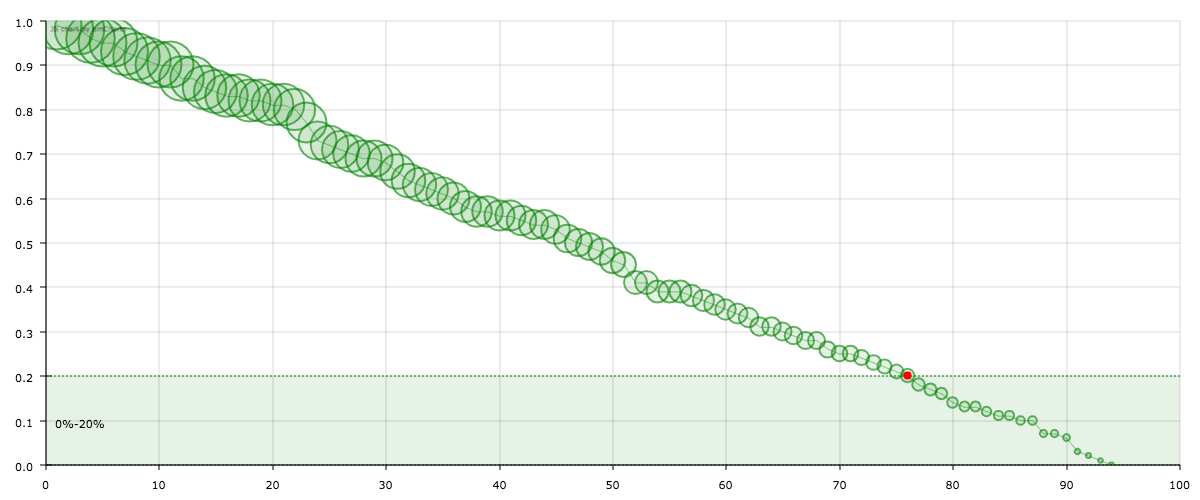}
\includegraphics[width=.5\textwidth]{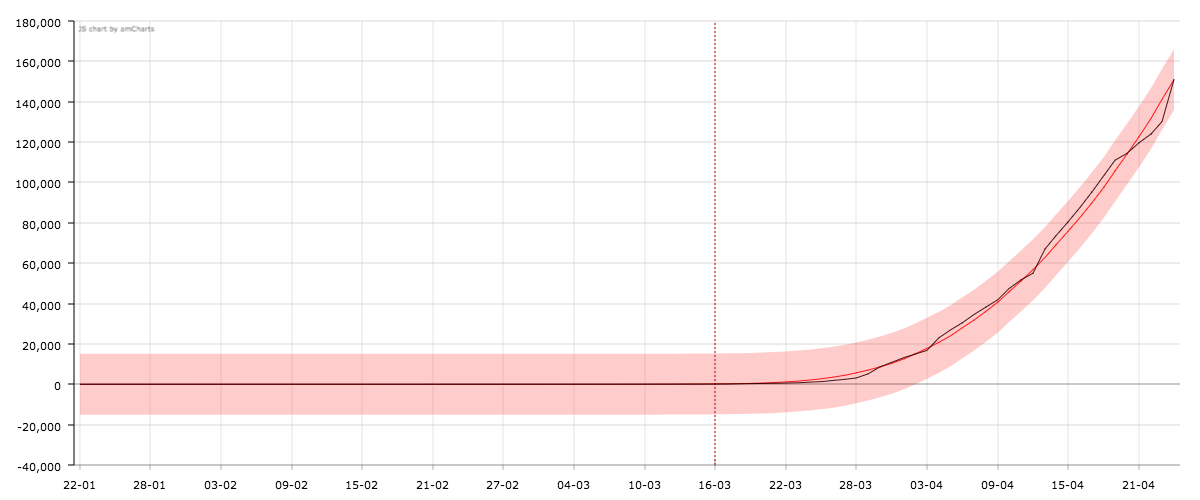}
\end{minipage}
\begin{minipage}{.99\textwidth}
\includegraphics[width=.5\textwidth]{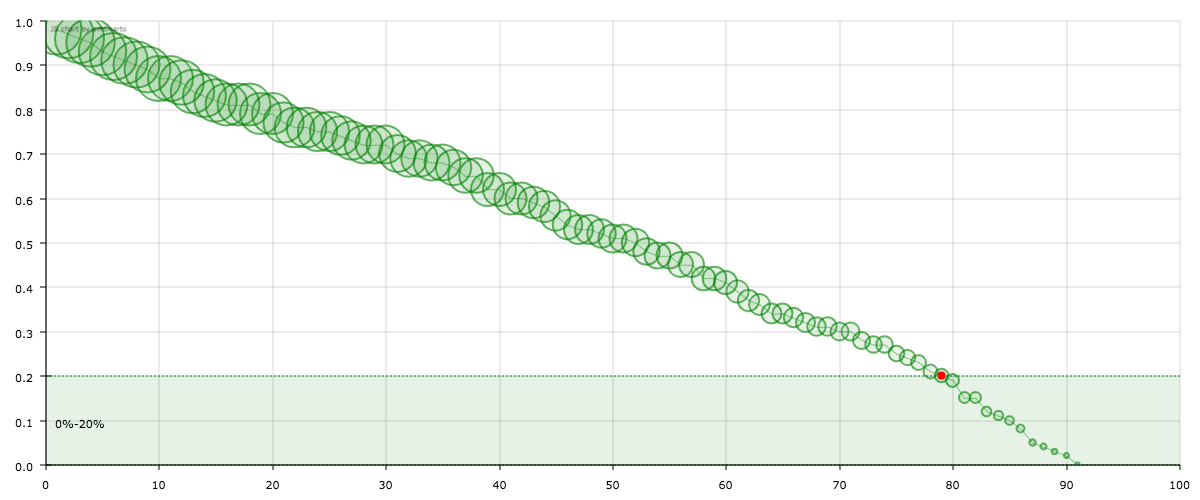}
\includegraphics[width=.5\textwidth]{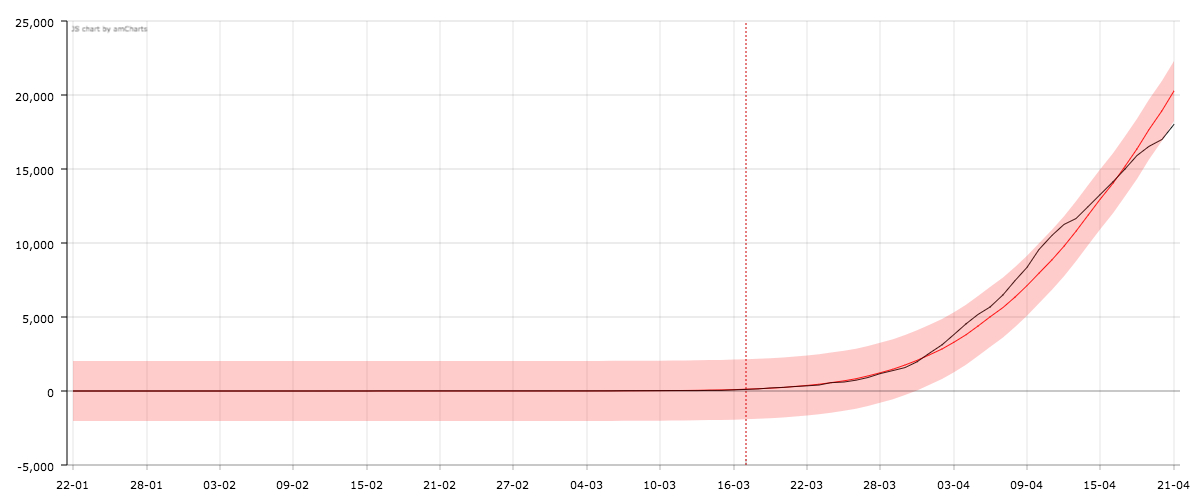}
\end{minipage}
\begin{minipage}{.99\textwidth}
\includegraphics[width=.5\textwidth]{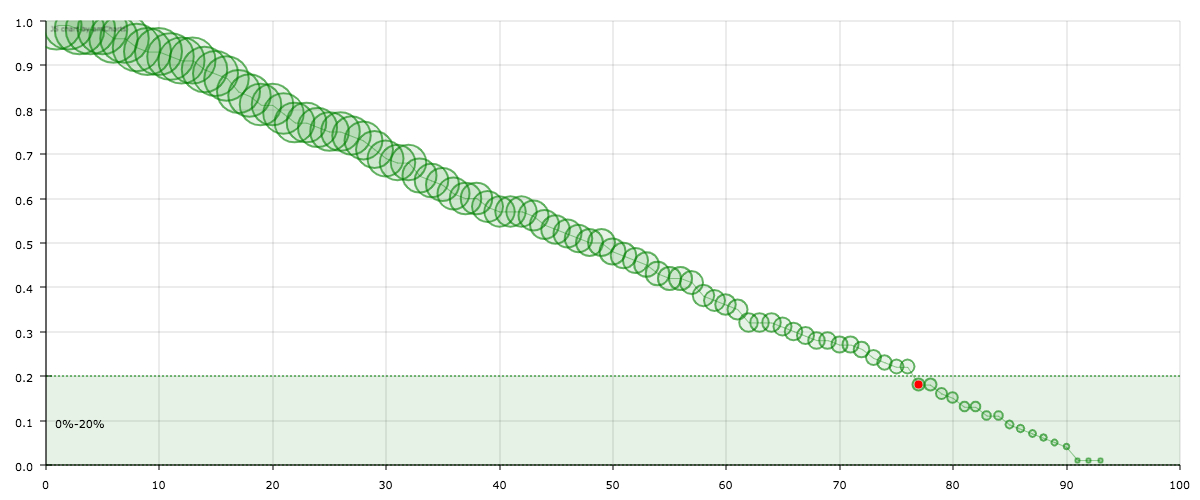}
\includegraphics[width=.5\textwidth]{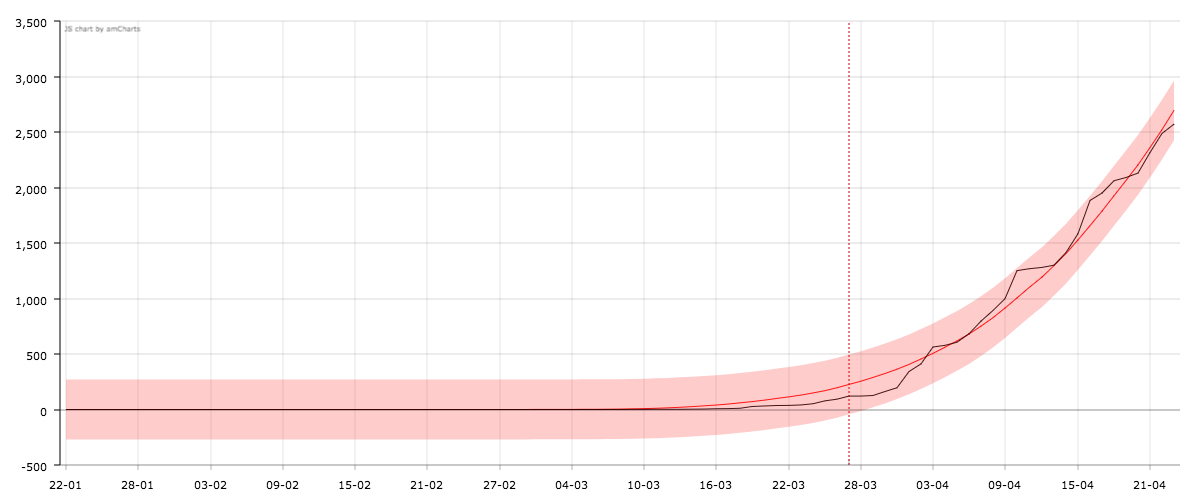}
\end{minipage}\\[2mm]
\caption{On the right, curve fitting of the sum of  the accumulated number of recovered and dead people in USA (top), United Kingdom (center) and Sweden (bottom): black line is the real value ($\mathcal X$) and red line is the approximation ($\widehat{\mathcal X}$). Vertical line signs the 100 cases. On the left, the corresponding survival curve of the virus $\mathcal S$. The red point signs the number of days after which a standard individual has a probability of staying in the group of infected people smaller than $0.2$.   \label{fig:survivalUSAUKSW} }
\end{figure}
\end{center}

\begin{center}\scriptsize
\begin{figure}[H] 
\begin{minipage}{.99\textwidth}
\includegraphics[width=.5\textwidth]{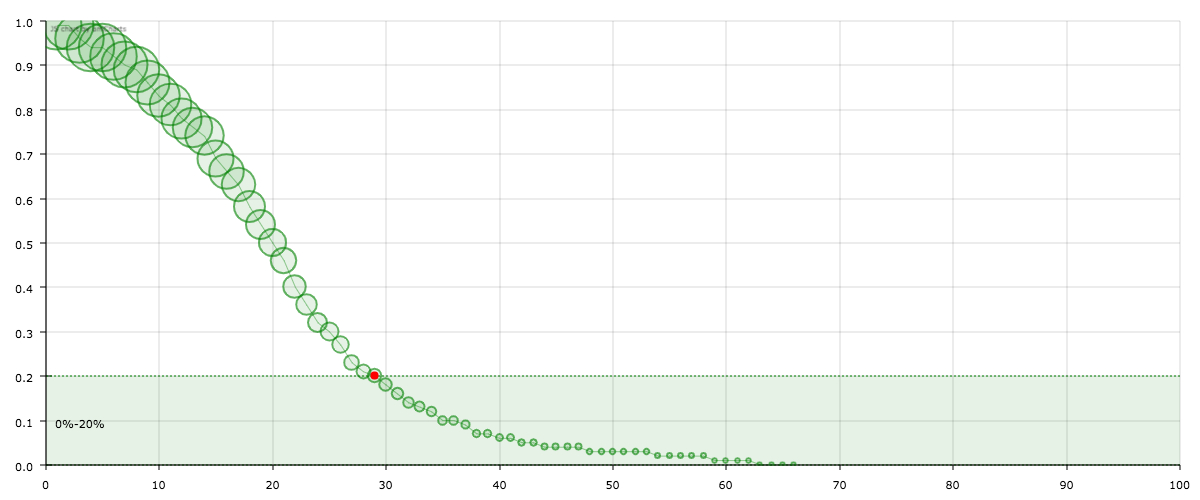}
\includegraphics[width=.5\textwidth]{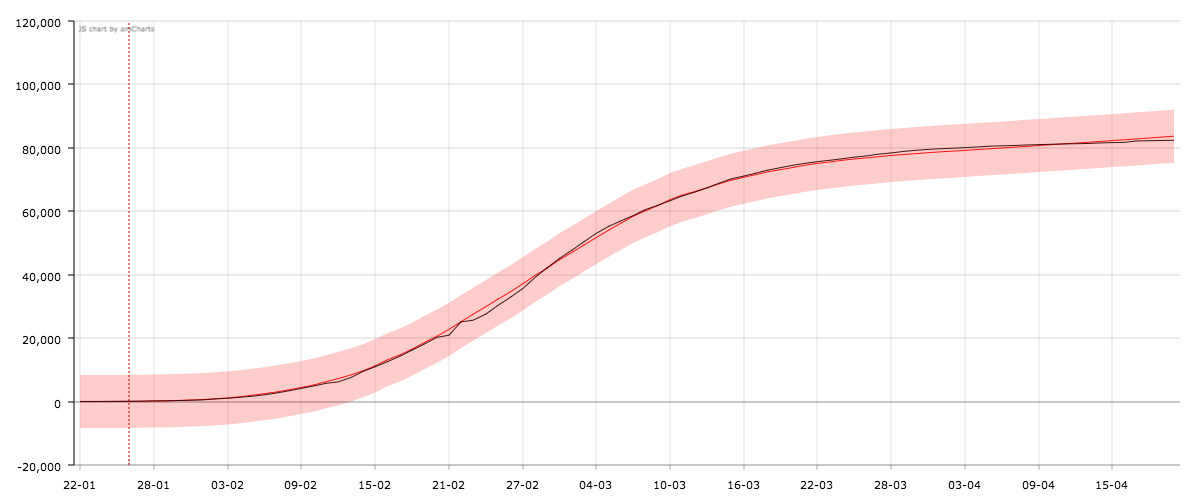}
\end{minipage}
\begin{minipage}{.99\textwidth}
\includegraphics[width=.5\textwidth]{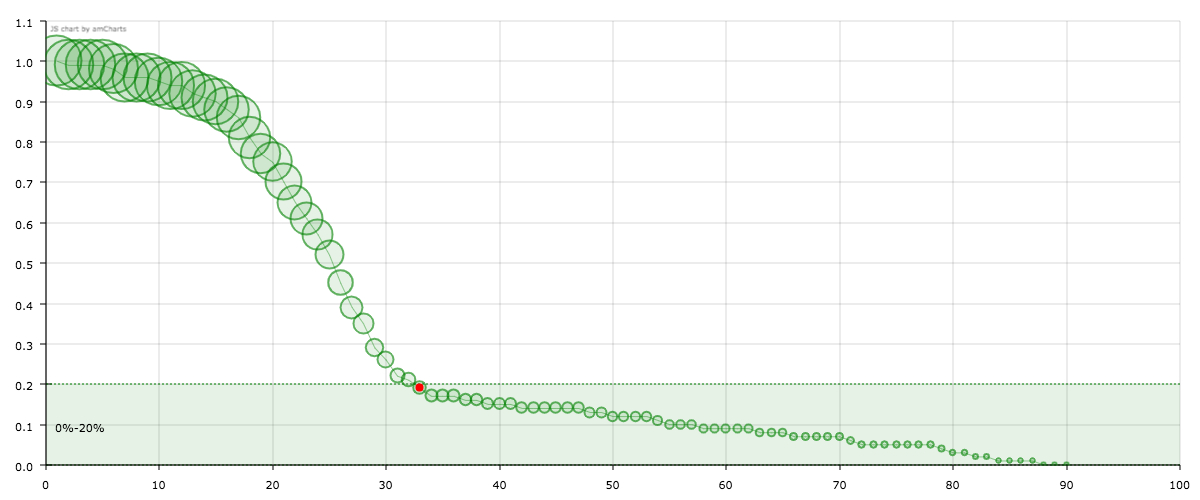}
\includegraphics[width=.5\textwidth]{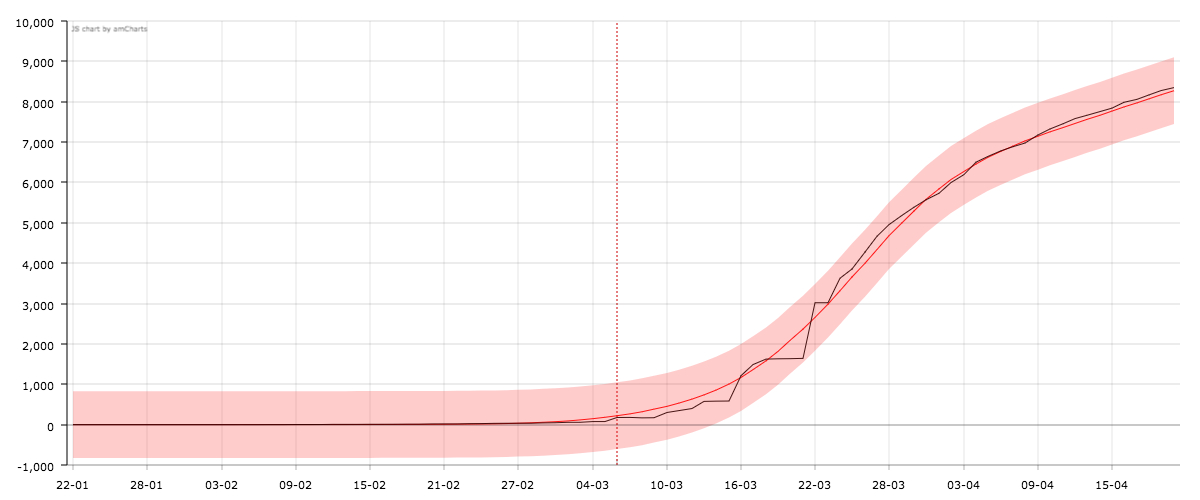}
\end{minipage}
\begin{minipage}{.99\textwidth}
\includegraphics[width=.5\textwidth]{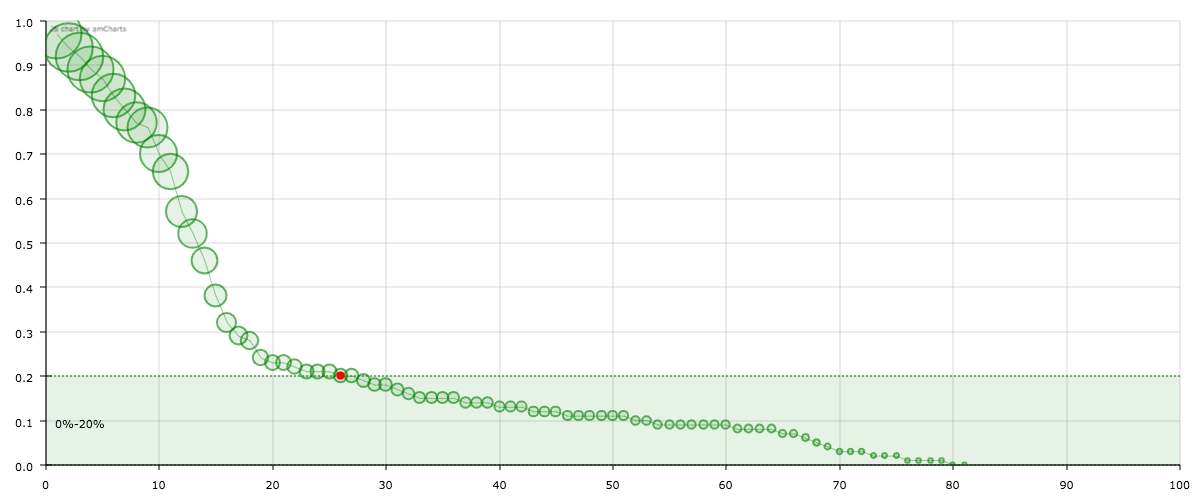}
\includegraphics[width=.5\textwidth]{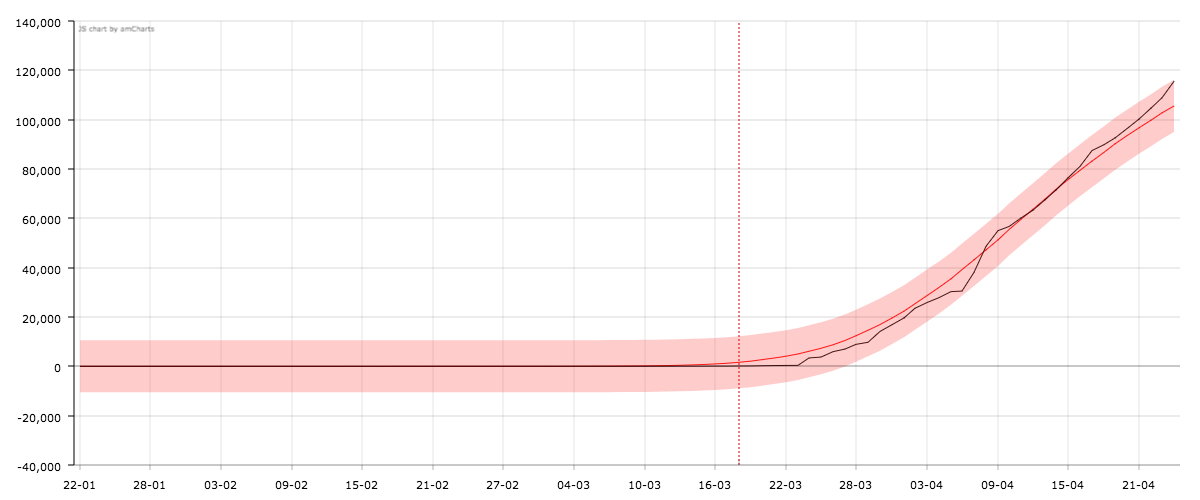}
\end{minipage}\\[2mm]
\caption{On the right, curve fitting of the sum of  the accumulated number of recovered and dead people in China (top), South Korea (center) and Germany (bottom): black line is the real value ($\mathcal X$) and red line is the approximation ($\widehat{\mathcal X}$). Vertical line signs the 100 cases. On the left, the corresponding survival curve of the virus $\mathcal S$. The red point signs the number of days after which a standard individual has a probability of staying in the group of infected people smaller than $0.2$.   \label{fig:survivalCHKOGE}}
\end{figure}
\end{center}
\newpage
\begin{center}\scriptsize
\begin{figure}[H] 
\begin{minipage}{.99\textwidth}
\includegraphics[width=.5\textwidth]{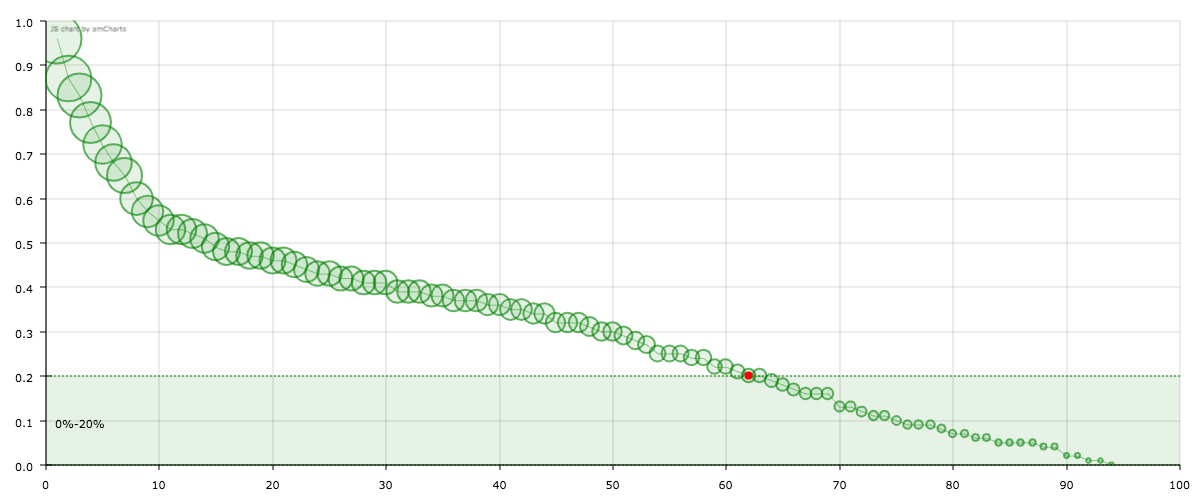}
\includegraphics[width=.5\textwidth]{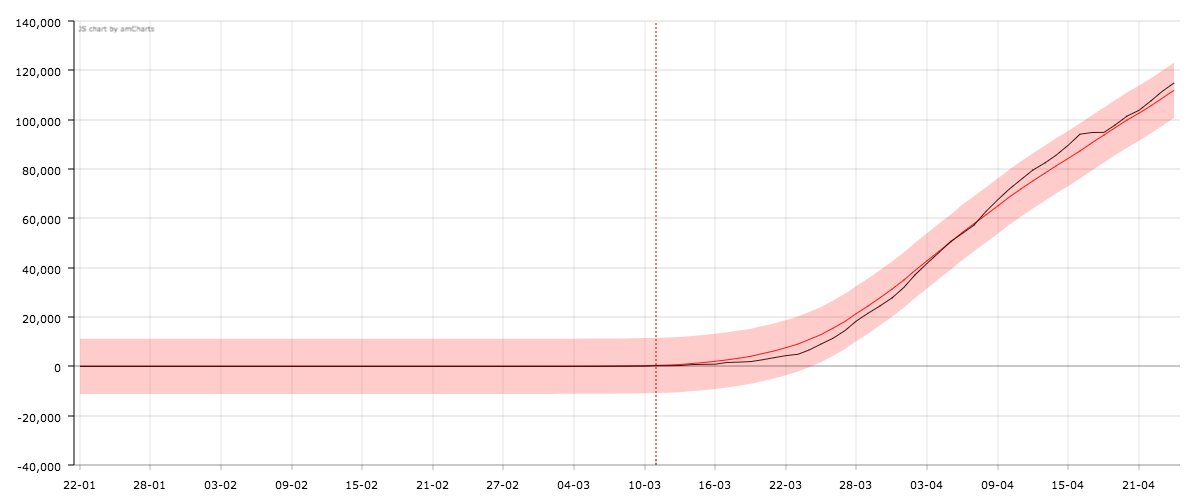}
\end{minipage}
\begin{minipage}{.99\textwidth}
\includegraphics[width=.5\textwidth]{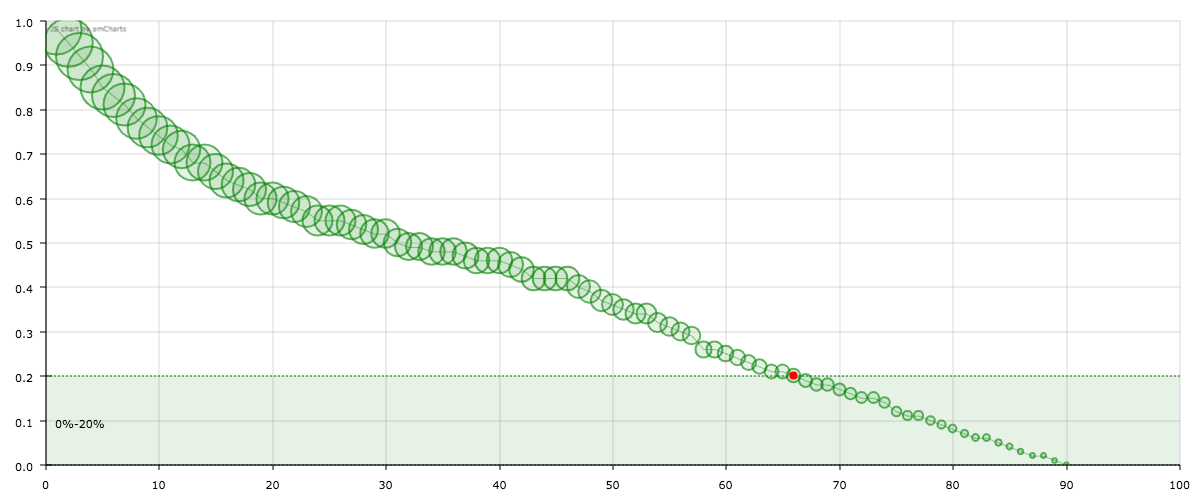}
\includegraphics[width=.5\textwidth]{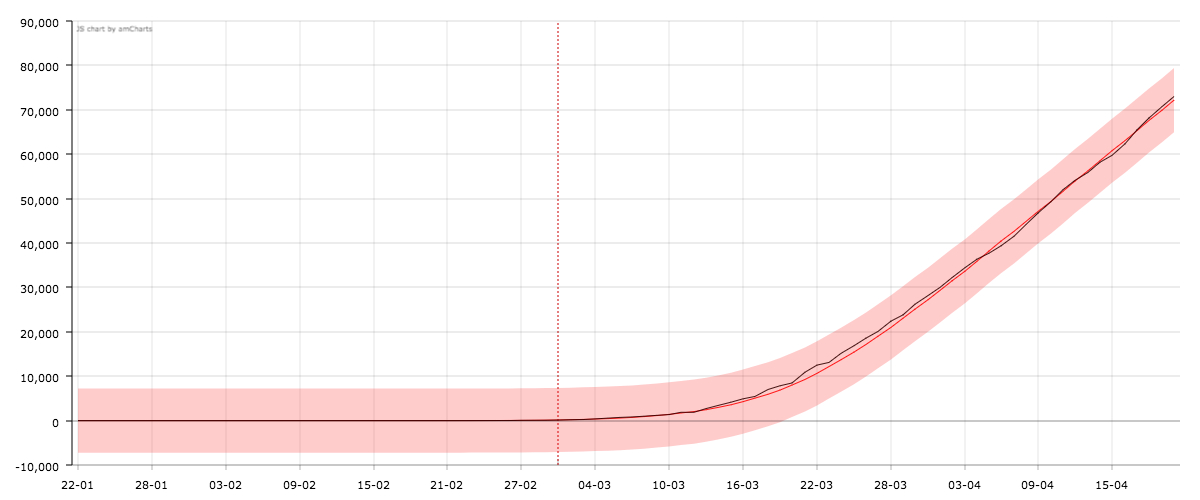}
\end{minipage}
\begin{minipage}{.99\textwidth}
\includegraphics[width=.5\textwidth]{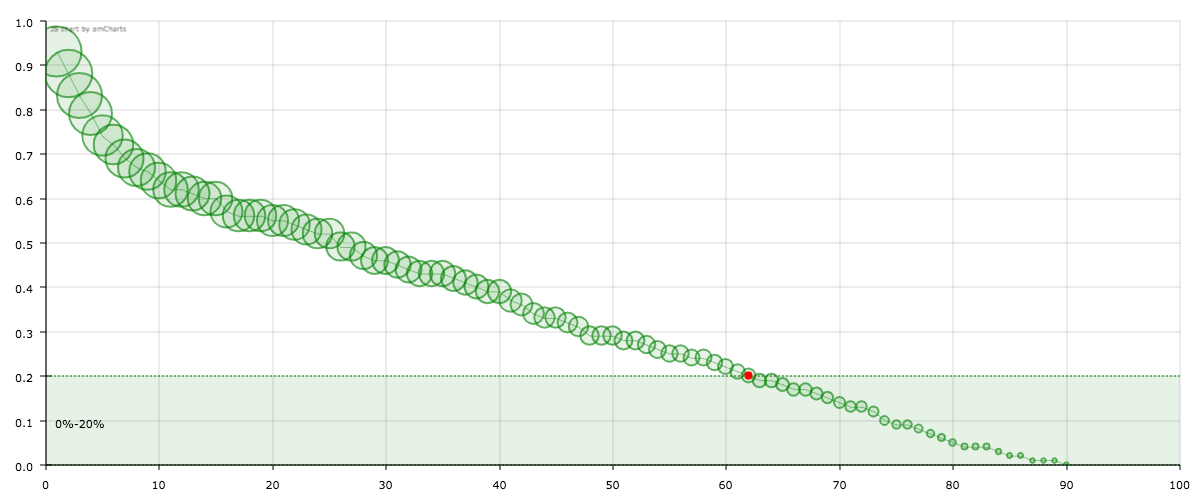}
\includegraphics[width=.5\textwidth]{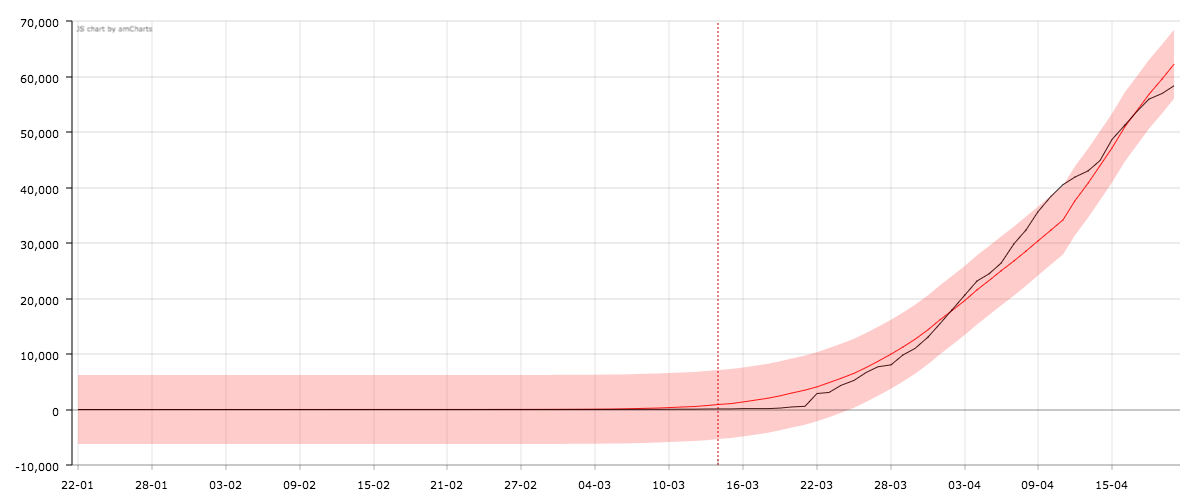}
\end{minipage}\\[2mm]
\caption{On the right, curve fitting of the sum of  the accumulated number of recovered and dead people in Spain (top), Italy (center) and France (bottom): black line is the real value ($\mathcal X$) and red line is the approximation ($\widehat{\mathcal X}$).   Vertical line signs the 100 cases. On the left, the corresponding survival curve of the virus $\mathcal S$. The red point signs the number of days after which a standard individual has a probability of staying in the group of infected people smaller than $0.2$.  \label{fig:survivalSPITFR} } 
\end{figure}
\end{center}

\begin{center}\scriptsize
\begin{figure}[H] 

\centering  {Survival curve of the virus $\mathcal S$ corresponding to the block formed by  USA, United Kingdom and Sweden.}\\[2mm]

\begin{minipage}{.66\textwidth}
\includegraphics[width=\textwidth]{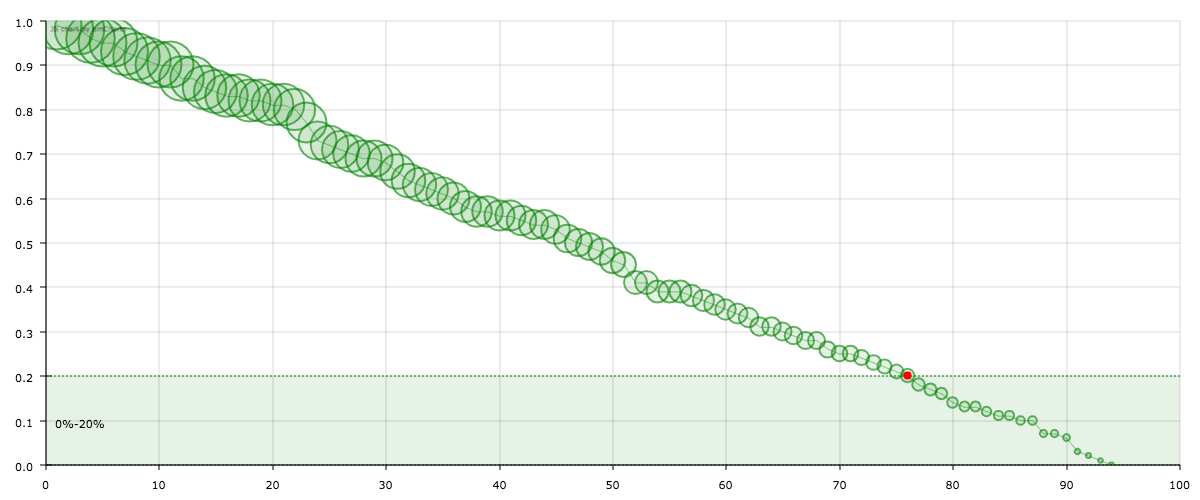}
\end{minipage}
\begin{minipage}{.33\textwidth}
\includegraphics[width=\textwidth]{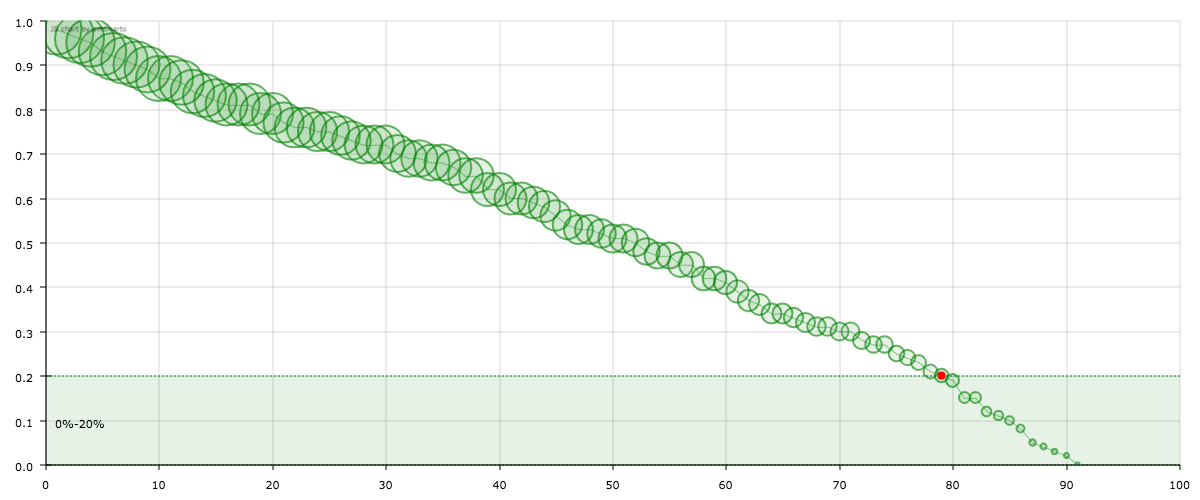}
\includegraphics[width=\textwidth]{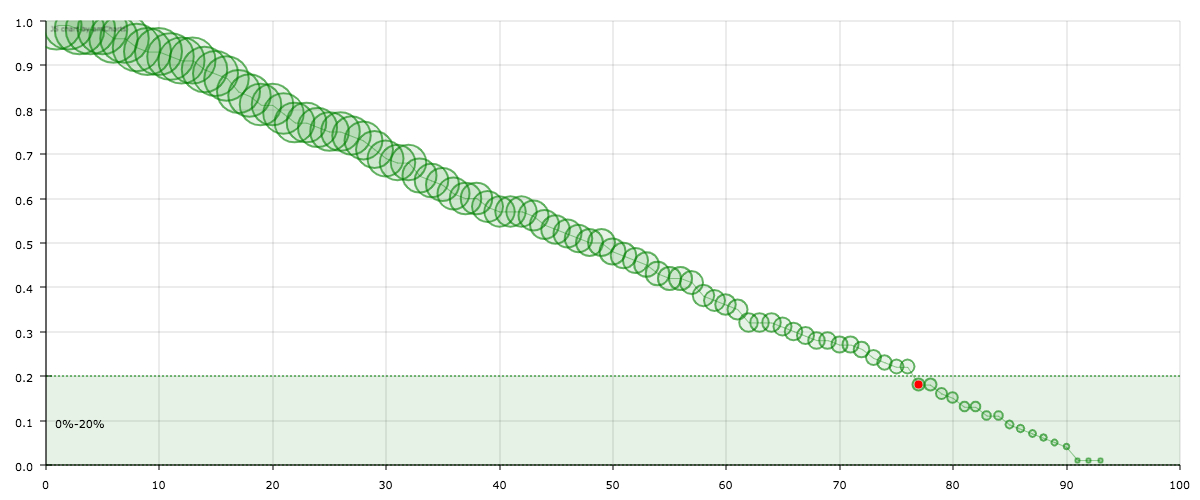}
\end{minipage}
\begin{minipage}{.32\textwidth}

\end{minipage}\\[2mm]

\centering {Survival curve of the virus $\mathcal S$ corresponding to the block formed by China, South Korea and Germany.}\\[2mm]

\begin{minipage}{.66\textwidth}
\includegraphics[width=\textwidth]{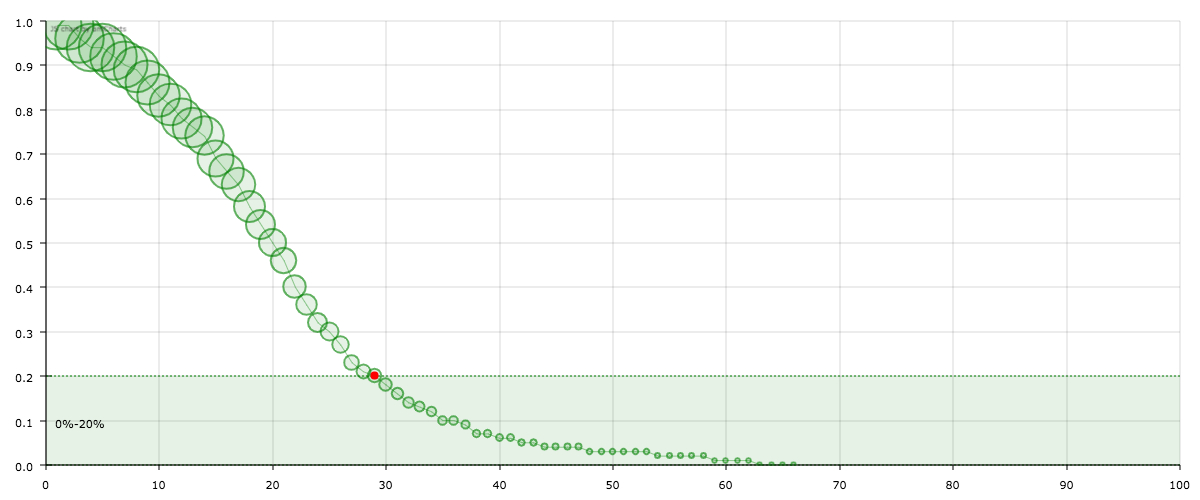}
\end{minipage}
\begin{minipage}{.33\textwidth}
\includegraphics[width=\textwidth]{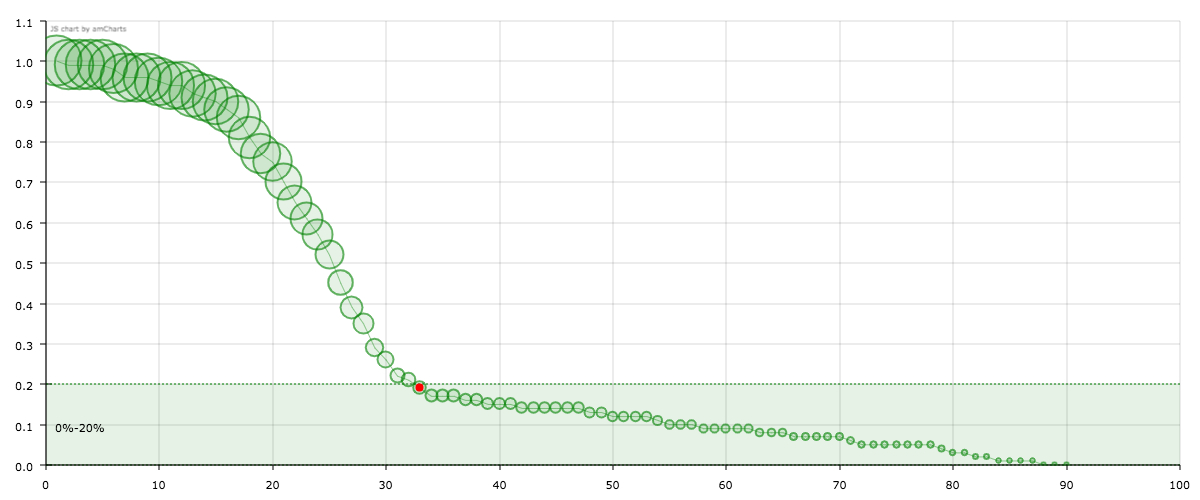}
\includegraphics[width=\textwidth]{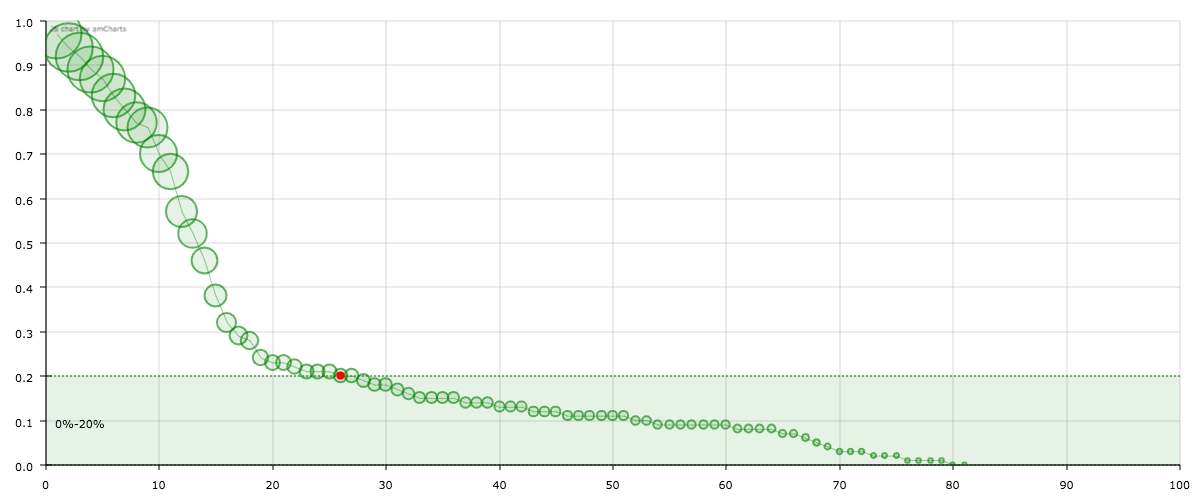}
\end{minipage}\\[2mm]

\centering {Survival curve of the virus $\mathcal S$ corresponding to the block formed by  Spain, Italy and France.} \\[3mm]

\begin{minipage}{.66\textwidth}
\includegraphics[width=\textwidth]{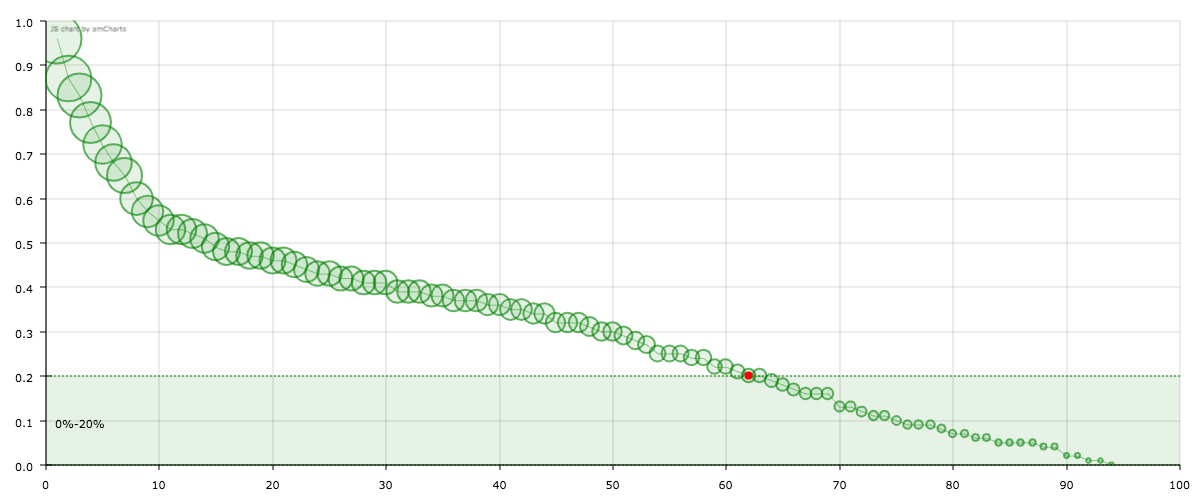}
\end{minipage}
\begin{minipage}{.33\textwidth}
\includegraphics[width=\textwidth]{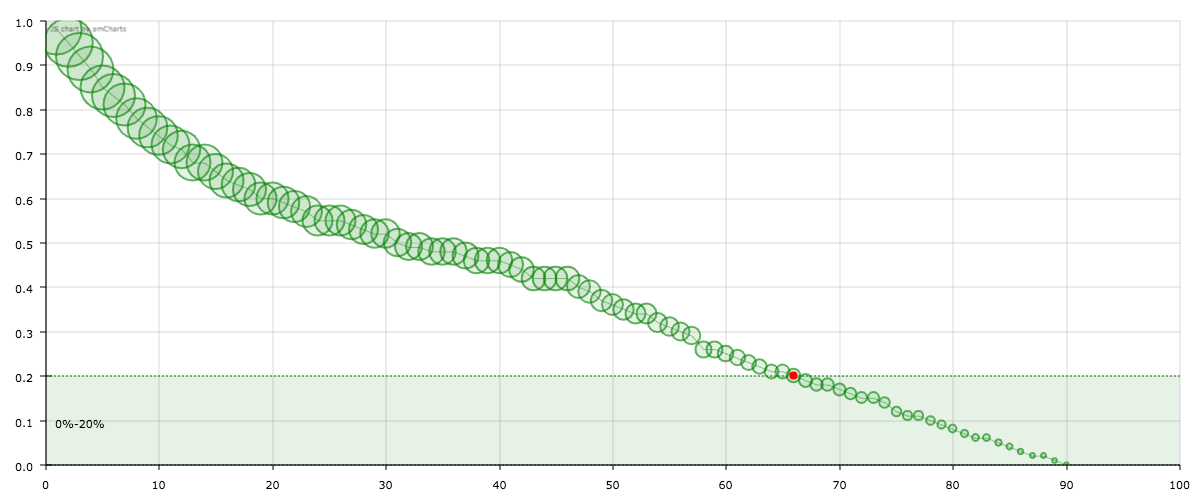}
\includegraphics[width=\textwidth]{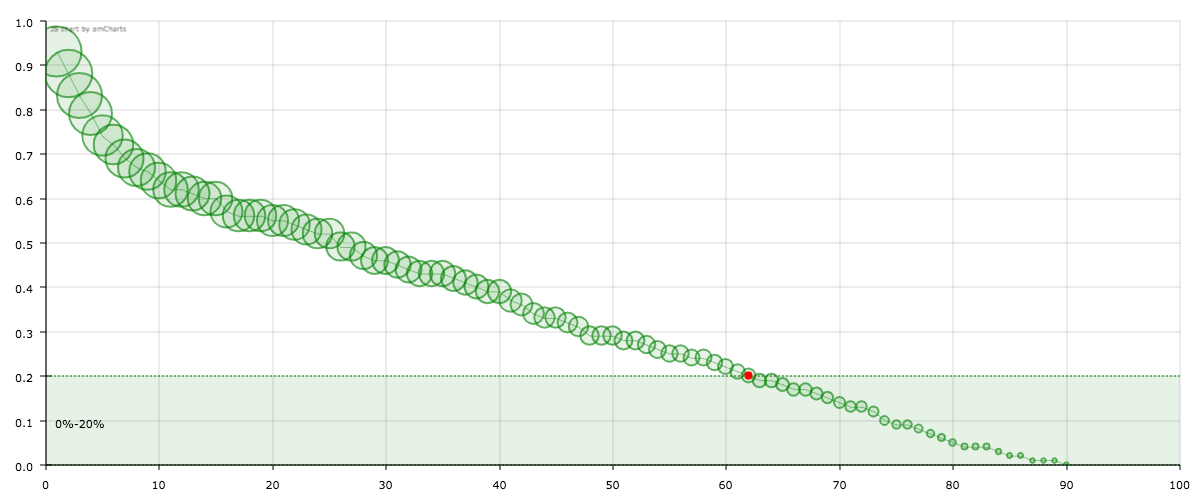}
\end{minipage}\\[2mm]
\caption{Pattern of the survival curves of the virus, $\mathcal S$,  for different blocks of countries. \label{fig:survivalpattern} } 
\end{figure}
\end{center}

\begin{center}\scriptsize
\begin{figure}[H]  
\begin{minipage}{.49\textwidth}
\includegraphics[width=\textwidth]{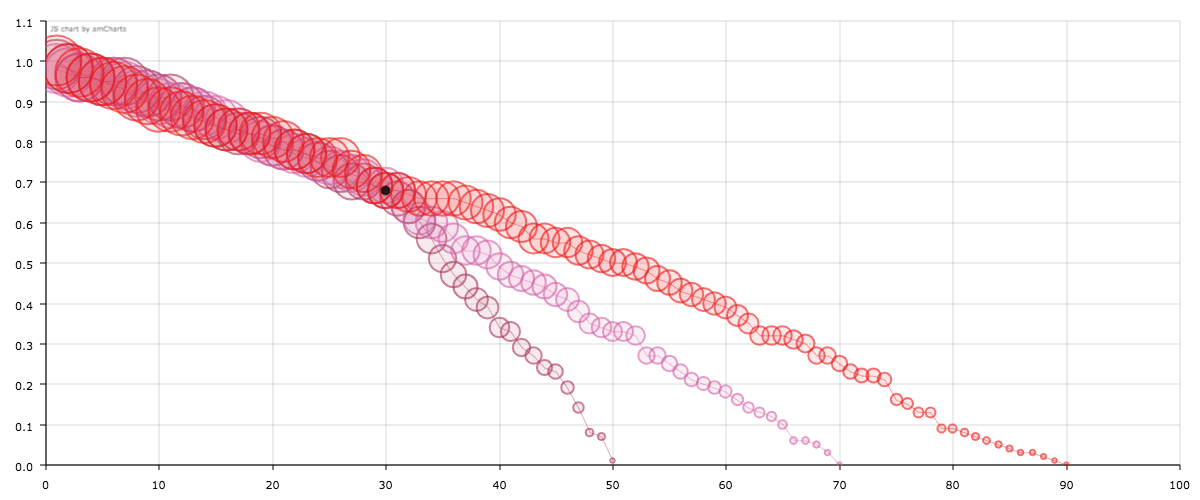}
\end{minipage}
\begin{minipage}{.49\textwidth}
\includegraphics[width=\textwidth]{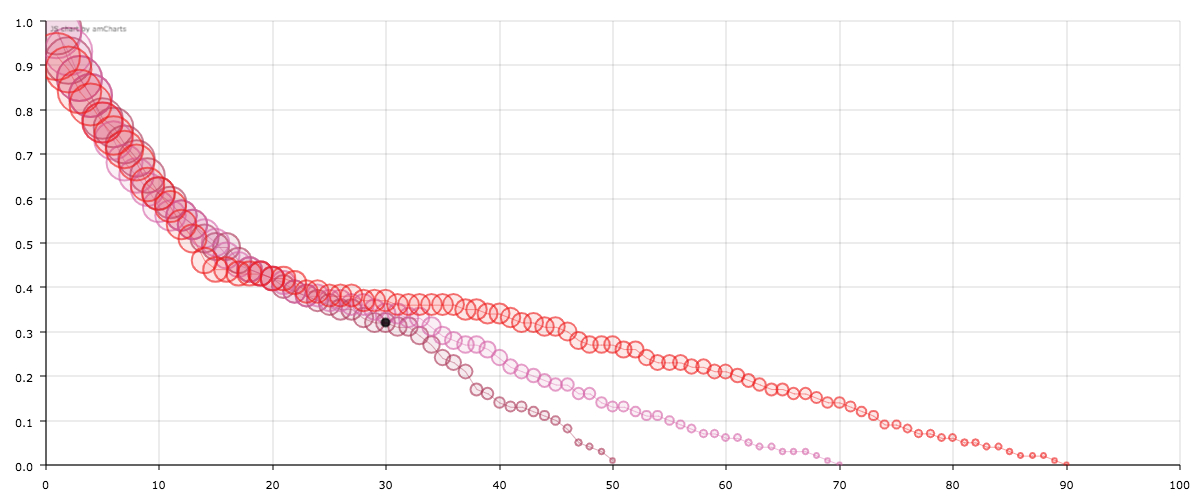}
\end{minipage}
\begin{minipage}{.49\textwidth}
\includegraphics[width=\textwidth]{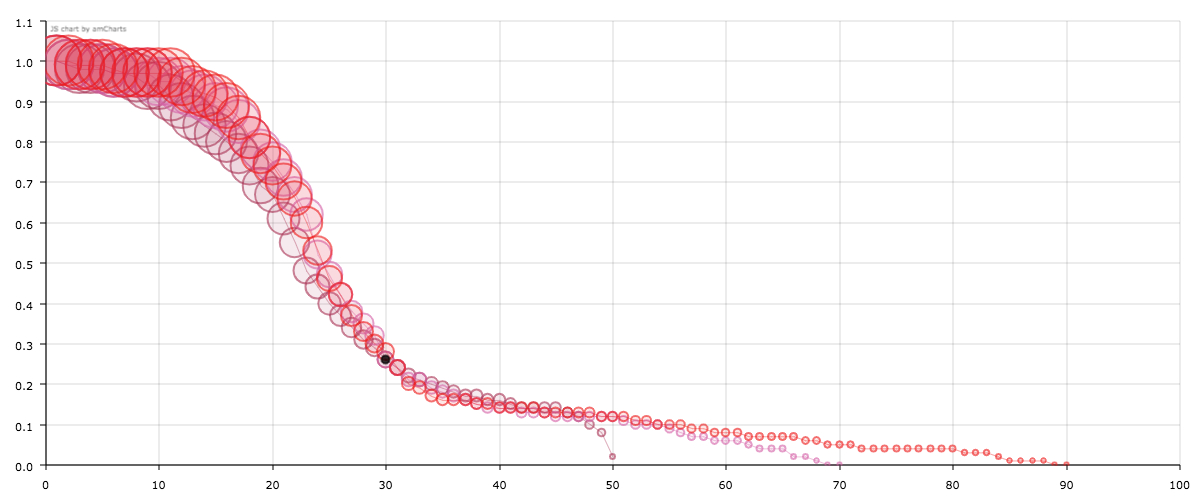}
\end{minipage}\\[2mm]
\caption{Survival curve of the virus considering different number of days for USA (left-top), Spain (right-top) and South Korea (center-bottom).
 Red circles corresponds to data of the first 90 days, pink circles to only the first 70 days and maroon circles to only the first 50 days.  
\label{fig:survivalpattern507090_USA_Korea_Spain}}
\end{figure}
\end{center}

\end{document}